# *Superluminal Spacetime Crystals Induced by Anomalous Velocity Modulation*

*Filipa R. Prudêncio*[(1,2)*], *Mário G. Silveirinha*[(1)]

[(1)]*University of Lisbon–Instituto Superior Técnico and Instituto de Telecomunicações, Avenida Rovisco Pais, 1, 1049-001 Lisboa, Portugal*

[(2)]*Instituto Universitário de Lisboa (ISCTE-IUL), Avenida das Forças Armadas 376, 1600-077 Lisbon, Portugal*

**Abstract**

Time-modulated media offer powerful opportunities for controlling light, yet extending such concepts to optical frequencies has remained challenging. Here we propose a different route to photonic spacetime crystals based on modulation of the anomalous velocity in low-symmetry conductors, particularly Weyl semimetals. We show that when driven by a strong optical pump, the anomalous velocity of Bloch electrons induces an ultrafast spacetime modulation that propagates with a superluminal phase velocity relative to the dielectric background. This self-induced modulation enables unidirectional light transport below the optical gap and, near the epsilon-near-zero point, gives rise to collective parametric resonance and stimulated emission of volume plasmons. These findings identify Weyl semimetals as a promising platform for realizing optical spacetime crystals and open a pathway toward active and nonreciprocal photonic systems governed by quantum geometric effects.

---

[*] To whom correspondence should be addressed: E-mail: filipa.prudencio@lx.it.pt

# Main text

Time modulation has recently emerged as an attractive strategy to engineer nonreciprocity and gain in electromagnetic platforms [1-6]. At microwave frequencies, temporal modulation of the material response is commonly achieved using electronic circuits, for example, by employing varactors [7]. Extending similar concepts to optical and infrared frequencies has proven much more challenging, primarily because interesting dynamical effects arise only when the modulation period becomes comparable to the optical cycle of the signal itself.

At these higher frequencies, the standard approach relies on third-order nonlinearities ($\chi^{(3)}$) to imprint a temporal modulation of the refractive index through the intensity envelope of a strong pump field. In this context, transparent conducting oxides (TCOs), such as indium tin oxide (ITO), have emerged as particularly promising materials [8-15]. Near their epsilon-near-zero (ENZ) point [16], $\varepsilon \approx 0$, small variations in carrier density or effective mass can induce large modulations of the permittivity, enabling ultrafast control of light–matter interactions. Despite remarkable progress and demonstrations of modulation on timescales approaching a single optical cycle [14], a clear experimental realization of optical amplification in ENZ-based platforms remains elusive.

These challenges underscore a fundamental limitation of $\chi^{(3)}$-based approaches: the modulation is externally imposed through the pump intensity envelope, rather than emerging from the material's intrinsic electronic response. This naturally motivates the exploration of systems with built-in $\chi^{(2)}$ nonlinearities capable of producing ultrafast, field-driven modulation [17-19]. In this context, low-symmetry conductors such as Weyl semimetals offer an appealing alternative platform.

Weyl semimetals have attracted attention in both condensed matter physics and photonics due to their unconventional optical responses rooted in intrinsic quantum geometry. These time-reversal–invariant materials lack inversion symmetry and therefore possess a nontrivial Berry curvature, which encapsulates the geometrical twisting of Bloch electron phases in momentum space. The Berry curvature gives rise to an additional contribution to the semiclassical velocity, known as the anomalous velocity, which causes electron wave packets to bend in real space under an applied electric field [20, 21].

Remarkably, this geometric correction has been shown to underlie a variety of exotic transport and optical phenomena, including the nonlinear Hall effect [22], photovoltaic rectification [23, 24], nonreciprocal electro-optic responses and optical gain [25-27], and even topological effects [28]. Furthermore, the unique chiral electronic structure of Weyl semimetals has been linked to ultra-giant second-order nonlinearities ($\chi^{(2)}$), several orders of magnitude larger than those found in conventional nonlinear materials [29-31]. These extraordinarily large nonlinearities arise from the metallic nature of the carriers: because electrons in such systems are relatively free, their displacement over an optical cycle can be much larger than that of bound electrons in dielectric crystals. For example, the nonlinear susceptibilities reported for Weyl semimetals can exceed those of GaAs, the established workhorse for $\chi^{(2)}$ nonlinear optics, by an order of magnitude [29]. On the other hand, a drawback of conducting materials compared to dielectric counterparts is their stronger dissipation, which can limit the net efficiency of nonlinear processes.

Building on these advances, we propose that low-symmetry conductors, such as Weyl semimetals, provide ideal platforms for realizing time-dependent optical responses, thereby enabling the implementation of photonic spacetime crystals at optical frequencies. We show that when these materials are driven by a strong optical pump, their response becomes intrinsically time dependent, even under a constant pump intensity. Because the underlying $\chi^{(2)}$ nonlinearity originates from the anomalous velocity, an inherently electronic mechanism, the resulting modulation is ultrafast by nature. This modulation of the anomalous velocity generates a spacetime crystal with a superluminal phase velocity relative to the dielectric background. The mechanism enables unidirectional light transport below the optical gap, offering a new pathway toward photonic nonreciprocity. Furthermore, when the pump frequency approaches twice the effective plasma frequency (the ENZ point), the induced spacetime modulation can efficiently excite volume plasmons, opening a route to optically driven amplification in ENZ-like Weyl systems.

We model the nonlinear optical response of a low-symmetry metal using the approach introduced in Ref. [32]. In this framework, the Bloch electrons near the Fermi level are effectively described by the transport equation $\frac{\partial \mathbf{p}}{\partial t} + \Gamma \mathbf{p} = -e\mathbf{E}$, where $\mathbf{p}$ denotes the quasi-momentum of the Bloch electrons, $\Gamma$ the collision frequency, $e > 0$ the elementary charge, and $\mathbf{E}$ the electric field. As usual, the free-electron current couples to Maxwell's equations through the current density $\mathbf{j} = -en\mathbf{v}$, with $n$ the electron density and $\mathbf{v}$ the electron velocity.

For conventional conductors, the relation between velocity and quasi-momentum is governed by the effective mass $m^*$, such that $\mathbf{v} = \mathbf{p}/m^*$. However, in low-symmetry conductors the velocity of each Bloch electron acquires an additional term, the anomalous velocity, arising from the geometric properties of the electronic wave functions in momentum space, encapsulated by the Berry curvature [20, 21]. This contribution can be incorporated phenomenologically by writing $\mathbf{v} = \frac{1}{m^*}\left[\mathbf{p} - \frac{m^* e}{\hbar^2 n}\left(\overline{\mathbf{D}}^{\mathrm{T}} \cdot \mathbf{p}\right) \times \mathbf{E}\right]$, where the second term represents a weighted anomalous velocity from electrons near the Fermi surface [32]. Here, $\hbar$ is the reduced Planck constant, the superscript "T" denotes matrix transpose, and $\overline{\mathbf{D}}$ is the dimensionless Berry curvature dipole of the material.

Crucially, this second term introduces a nonlinearity, as it depends on the product of the quasi-momentum and the electric field, and it is precisely this nonlinear contribution that underpins the physics explored in this Letter. It was shown in Ref. [32] that this effective description reproduces the results of the semiclassical Boltzmann theory [22, 25]. As shown next, this effective formalism is particularly powerful, as it provides a transparent and compact means to characterize the optical response of low-symmetry conductors under strong optical pumping, without the need to solve the full microscopic transport equations.

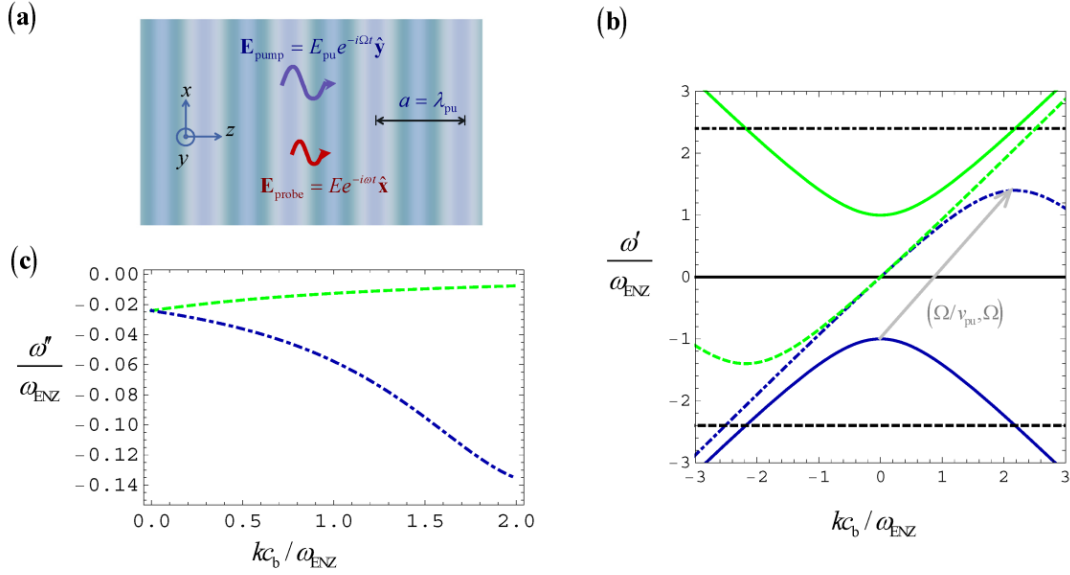

**Fig. 1 (a)** Schematic of the spacetime crystal generated by the optical pump propagating through a Weyl semimetal. **(b)** Dispersion of the Bloch modes in the spacetime crystal for a weak modulation and $\Omega = 2.4\omega_{\mathrm{ENZ}} = 2\pi \times 30\,\mathrm{THz}$. The solid lines show the dispersion of the unmodulated (undriven) material. The blue dash–dotted line (green dashed line) corresponds to the sideband $l=1$ ($l=-1$) generated from the negative (positive) frequency branch by a shift along the spacetime lattice vector $\left(\Omega/v_{\mathrm{pu}}, \Omega\right)$ (gray arrow). The sidebands associated with the magnetostatic mode (in black) are also shown. **(c)** Damping factor $\omega''$ of the two Bloch modes lying within the spectral gap of the unmodulated crystal, calculated for a modulation depth $\delta_M = 0.005$. The color code is the same as in (b).

In our setup, we assume that the optical pump has a frequency $\Omega$ different from that of the probe signal, $\omega$, and that it propagates within a transparency window of the material. Furthermore, the pump intensity is taken to be much larger than that of the probe, and both waves are assumed to propagate along the $z$-axis (see Fig. 1a). Under these conditions, the linearized effective velocity of the electrons relevant to the probe dynamics can be expressed as $\mathbf{v} = \dfrac{1}{m^*}\left[\mathbf{p} - \dfrac{m^*e}{\hbar^2 n}\left(\overline{\mathbf{D}}^{\mathrm{T}}\cdot\mathbf{p}_{\mathrm{pu}}\right)\times\mathbf{E} - \dfrac{m^*e}{\hbar^2 n}\left(\overline{\mathbf{D}}^{\mathrm{T}}\cdot\mathbf{p}\right)\times\mathbf{E}_{\mathrm{pu}}\right]$. Here, $\mathbf{E}_{pu}$ and $\mathbf{p}_{pu}$ denote the electric field and quasi-momentum of the Bloch electrons associated with the strong optical pump, respectively, linked by $\frac{\partial}{\partial t}\mathbf{p}_{pu} + \Gamma \mathbf{p}_{pu} = -e\mathbf{E}_{pu}$. For clarity, all fields related to the pump are labeled with the subscript "pu", whereas fields associated with the probe signal carry no subscript. Using $\mathbf{j} = -en\mathbf{v}$, we find that the linearized electrodynamics of the weak probe signal is governed by,

$$\nabla \times \mathbf{H} = \mathbf{j} + \varepsilon_b \partial_t \mathbf{E}, \qquad \nabla \times \mathbf{E} = -\mu_b \partial_t \mathbf{H}, \tag{1a}$$

$$\mathbf{j} = \varepsilon_b \omega_{ENZ}^2 \left[ \tilde{\mathbf{p}} - \frac{1}{E_{0n}}\left(\overline{\mathbf{D}}^{\mathrm{T}} \cdot \tilde{\mathbf{p}}_{pu}\right) \times \mathbf{E} - \frac{1}{E_{0n}}\left(\overline{\mathbf{D}}^{\mathrm{T}} \cdot \tilde{\mathbf{p}}\right) \times \mathbf{E}_{pu} \right], \tag{1b}$$

$$\frac{\partial \tilde{\mathbf{p}}}{\partial t} + \Gamma \tilde{\mathbf{p}} = \mathbf{E}, \tag{1c}$$

where we have normalized the quasi-momentum of both the probe and the pump as $\tilde{\mathbf{p}} = \mathbf{p}/(-e)$ and $\tilde{\mathbf{p}}_{pu} = \mathbf{p}_{pu}/(-e)$. Here, $\varepsilon_b, \mu_b$ are the permittivity and permeability of the background dielectric and $E_{0n} = \frac{n\hbar^2}{m^* e}$. We introduced $\omega_{ENZ} = \sqrt{\frac{e^2 n}{m^* \varepsilon_b}}$, corresponding to the ENZ point of the linear response. The constant $E_{0n}$ has units of electric field and can be directly expressed in terms of the ENZ frequency as $E_{0n} = \omega_{ENZ}^2 \frac{\varepsilon_b \hbar^2}{e^3}$. For an ENZ frequency in the terahertz range ($\omega_{ENZ} \sim 2\pi \times 10\,\mathrm{THz}$), its magnitude is on the order of $\frac{\varepsilon_b}{\varepsilon_0} \times 10^5 V/m$, where $\varepsilon_0$ denotes the vacuum permittivity.

In essence, system (1) describes the electrodynamics of a time-varying plasma whose material response is modulated by the optical pump through anomalous velocity coupling. Notably, the modulation mechanism considered here is qualitatively distinct from that in TCOs: it does not rely on pump intensity in a $\chi^{(3)}$ medium, but instead on a $\chi^{(2)}$ nonlinearity originating from the anomalous velocity. This second-order response modulates the material properties at the pump frequency without requiring amplitude modulation by femtosecond pulses. Consequently, a nonlinear medium with broken inversion symmetry, when driven by a strong continuous pump, naturally behaves as a photonic (space–)time crystal, since a constant-intensity field directly induces modulation at the pump frequency [17, 18].

Next, we focus on materials belonging to the 3m point group, specifically the Weyl semimetal bismuth telluroiodide (BiTeI), with the crystal symmetry axis oriented along the $y$-direction. The corresponding Berry dipole tensor is $\overline{\mathbf{D}} = D(\hat{\mathbf{x}} \otimes \hat{\mathbf{z}} - \hat{\mathbf{z}} \otimes \hat{\mathbf{x}})$, with $D \sim 0.01$ for a broad range of chemical potentials [33, 34]. We model the optical response of BiTeI using parameters similar to those in Ref. [35]: $\varepsilon_\infty \sim 13$, $\omega_{\mathrm{ENZ}} \equiv \omega_{\mathrm{p}}/\sqrt{\varepsilon_\infty} \sim 2\pi \times 12.5\,\mathrm{THz}$ and $\Gamma \sim 2\pi \times 3.5\,\mathrm{THz}$. To ensure operation well below the interband threshold, we consider an electron concentration four times smaller than in Ref. [35]. The background permittivity is $\varepsilon_{\mathrm{b}} = \varepsilon_0 \varepsilon_\infty$ and the permeability is $\mu_{\mathrm{b}} = \mu_0$. BiTeI has a bandgap of 0.36 eV, ensuring that interband effects become relevant only above 90 THz.

We further assume that the optical pump is a plane wave polarized along $y$, such that

$$\mathbf{E}_{\mathrm{pu}} \approx \hat{\mathbf{y}} E_{0\mathrm{pu}} \cos\left(k_{\mathrm{pu}}\left(z - v_{\mathrm{pu}} t\right)\right) \quad \text{and} \quad \tilde{\mathbf{p}}_{\mathrm{pu}} \approx -\hat{\mathbf{y}} \frac{E_{0\mathrm{pu}}}{\Omega} \sin\left(k_{\mathrm{pu}}\left(z - v_{\mathrm{pu}} t\right)\right),$$

where $v_{\mathrm{pu}} = c_{\mathrm{b}} / \sqrt{1 - \left(\omega_{\mathrm{ENZ}}/\Omega\right)^2}$ is the pump phase velocity, $\Omega$ the pump frequency, $k_{\mathrm{pu}} = \dfrac{\Omega}{v_{\mathrm{pu}}}$ the pump wavenumber, and $c_{\mathrm{b}} = 1/\sqrt{\varepsilon_{\mathrm{b}} \mu_{\mathrm{b}}}$ the speed of light in the background dielectric.

Owing to the dispersive Drude-type nature of the metallic response, the pump phase velocity is superluminal with respect to the background ($v_{pu} > c_b$). For simplicity, dissipative effects on the pump are neglected, an approximation justified when $\Omega \gg \Gamma$.

In the first example, we consider the case where the probe signal is a transverse electromagnetic wave polarized along $x$, so that $\mathbf{E} = E(z,t)\hat{\mathbf{x}}$, $\mathbf{H} = H(z,t)\hat{\mathbf{y}}$ and $\tilde{\mathbf{p}} = \tilde{p}(z,t)\hat{\mathbf{x}}$. Under these conditions, system (1) reduces to

$$\frac{\partial H}{\partial z} = -\varepsilon_b \omega_{ENZ}^2 (z - v_{pu}t) \tilde{p} - \varepsilon_b \frac{\partial E}{\partial t}, \qquad \frac{\partial E}{\partial z} = -\mu_b \frac{\partial H}{\partial t}, \qquad (2a)$$

$$\frac{\partial \tilde{p}}{\partial t} + \Gamma \tilde{p} = E, \qquad (2b)$$

where $\omega_{ENZ}^2(z') = \omega_{ENZ}^2 \left[1 + \delta_M \cos(k_{pu} z')\right]$, with $\delta_M = D \frac{E_{0pu}}{E_{0n}}$ the modulation depth. From the structure of the above equations, it is evident that the pump field effectively induces a modulation of the ENZ frequency: $\omega_{ENZ} = \omega_{ENZ}(z,t)$. Because the pump field is a traveling wave, its spatial and temporal dependencies are locked through its phase velocity, such that the ENZ frequency depends only on $z' = z - v_{pu}t$. Consequently, the pump establishes a traveling-wave spacetime crystal within the Weyl semimetal. Traveling-wave spacetime crystals have been extensively studied in the literature due to their conceptual simplicity [3, 36]. It has been shown that such systems can emulate the physics of moving media [37-40], enable the design of metamaterials with exotic magnetoelectric couplings [41-42], and host unconventional topological phases [43], amongst others [44-48].

The electromagnetic modes in the spacetime crystal are generalized Bloch waves of the form $u_k(z-v_{pu}t)e^{-i\omega_k t}e^{ikz}$, where $u_k(z)=u_k(z+a)$ is the cell-periodic amplitude, $k$ is the Bloch wavenumber, and $\omega$ is the Bloch frequency. Generally, the Bloch frequency is a complex function of $k$, $\omega_k = \omega'_k + i\omega''_k$ with $-\omega''_k$ representing the decay rate in time of the mode. The spatial period is determined by the pump wavelength ($a = \lambda_{pu} \equiv 2\pi/k_{pu} = v_{pu}2\pi/\Omega$).

For weak modulations, the spectrum of the spacetime crystal can be obtained from that of the unmodulated plasma by periodic band folding, which creates spectral replicas (sidebands) shifted in $(k,\omega)$-space along the directions $l(\Omega/v_{pu},\Omega)$, with $l = 0,\pm 1,\pm 2,...$ [47]. This construction is illustrated in Fig. 1b for a pump with $\Omega = 2.4\omega_{ENZ} = 2\pi \times 30 \text{THz}$. The solid lines represent the spectrum of the unmodulated plasma, consisting of two parabolic bands with positive and negative frequencies ($\omega \approx \pm\sqrt{\omega_{ENZ}^2 + c_b^2 k^2}$, shown in green and blue, respectively) and a flat, magnetostatic-like band at zero frequency ($\omega \approx 0$, black line). The dashed lines in Fig. 1b correspond to sidebands associated with $l = -1$, while the dash–dotted lines represent those associated with $l = +1$. Within first-order perturbation theory, a sinusoidal modulation only enables transitions associated with $l = \pm 1$ [15], so higher-order replicas can be neglected. Using methods similar to those in Refs. [41, 46], we numerically solved Eq. (2) to obtain the exact spectrum. For modulation depths up to $\delta_M = 0.25$, the exact results are visually indistinguishable from those predicted by the simple band-folding construction described above. For realistic pump intensities in continuous-wave operation, $E_{0pu} \sim 10^6 V/m$, the modulation depth is well within this range: $\delta_M \sim 0.005$.

The remarkable feature of the spectrum shown in Fig. 1b is that the replicas with $l = \pm 1$ effectively create a propagation channel within the bandgap of the unmodulated plasma ($\omega < \omega_{ENZ}$). The spacetime modulation enables light with a nominally forbidden frequency to enter, propagate through, and exit the crystal by coupling to its dynamically generated sidebands. Because the modulation travels with the pump, only co-propagating probe signals can couple into this channel, resulting in intrinsically unidirectional propagation, a key feature for realizing nonreciprocal photonic devices. Figure 1c shows the decay rate of the Bloch waves associated with this "forbidden-band" channel. As seen, the sideband $l = -1$ corresponding to the positive-frequency branch $\omega \approx +\sqrt{\omega_{ENZ}^2 + c_b^2 k^2}$ (green dashed line) experiences the weakest attenuation. The overall transmission efficiency through a spacetime crystal slab depends on the modulation depth $\delta_M$, though it is also influenced by interface effects, which are beyond the scope of the present analysis. Interestingly, due to the superluminal nature of the modulation, the replicas in Fig. 1b always intersect the $\omega = 0$ line at the $k = 0$ point, resulting in a quasi-linear dispersion, $\omega \approx \frac{c_b^2}{v_{pu}} k$, similar to that of the background dielectric as $v_{pu}$ is only marginally larger than $c_b$. This feature may facilitate efficient coupling between the spacetime crystal and external fields, further enhancing the potential for direction-selective light transport.

Next, we turn our attention to parametric amplification. As discussed in Ref. [47], this process can occur when replicas originating from a static negative-frequency branch intersect a static positive-frequency band. However, as shown in Fig. 1b, no intersections exist between the static positive-frequency band (green solid line) and the sideband of the negative-frequency branch (blue dash–dotted line). Consequently, this system does not support the parametric amplification of transverse waves. Nevertheless, as discussed below, the amplification of longitudinal modes, i.e., volume plasmons, remains possible in principle.

Volume plasmons are excitations characterized by a negligible magnetic field and an electric field oriented parallel to the propagation direction. In the static crystal, these modes correspond to flat bands at the ENZ point, $\omega = \pm\omega_{ENZ}$ (see the solid lines in Fig. 2a), with a dispersion that is independent of the wave vector. To examine how such excitations are modified by the spacetime drive, we substitute the ansatz $\mathbf{E} = E(z,t)\hat{\mathbf{z}}$, $\mathbf{H} = 0$, and $\tilde{\mathbf{p}} = \tilde{p}(z,t)\hat{\mathbf{z}}$ into system (1), assuming the same pump wave as before. It can then be shown that the dynamics of these longitudinal modes are governed by:

$$\frac{\partial E}{\partial t} = -\omega_{ENZ}^2(z-vt)\tilde{p}, \qquad \frac{\partial \tilde{p}}{\partial t} = E - \Gamma\tilde{p}, \qquad (3)$$

where the spacetime-varying ENZ frequency is defined as before. Interestingly, although the ENZ frequency varies in both space and time, the spatial dependence plays no dynamical role as $z$ can be treated as a fixed parameter. Consequently, the Bloch waves retain the same structure, $u_k(z-v_{pu}t)e^{-i\omega t}e^{ikz}$, with all parameters defined as in the transverse case. However, $\omega$ is now independent of the wave vector and is determined solely by the pump frequency, $\omega = \omega(\Omega)$. This behavior, which is intuitively consistent with the band-folding picture discussed earlier (see Fig. 2a), remains valid even for strong modulation amplitudes.

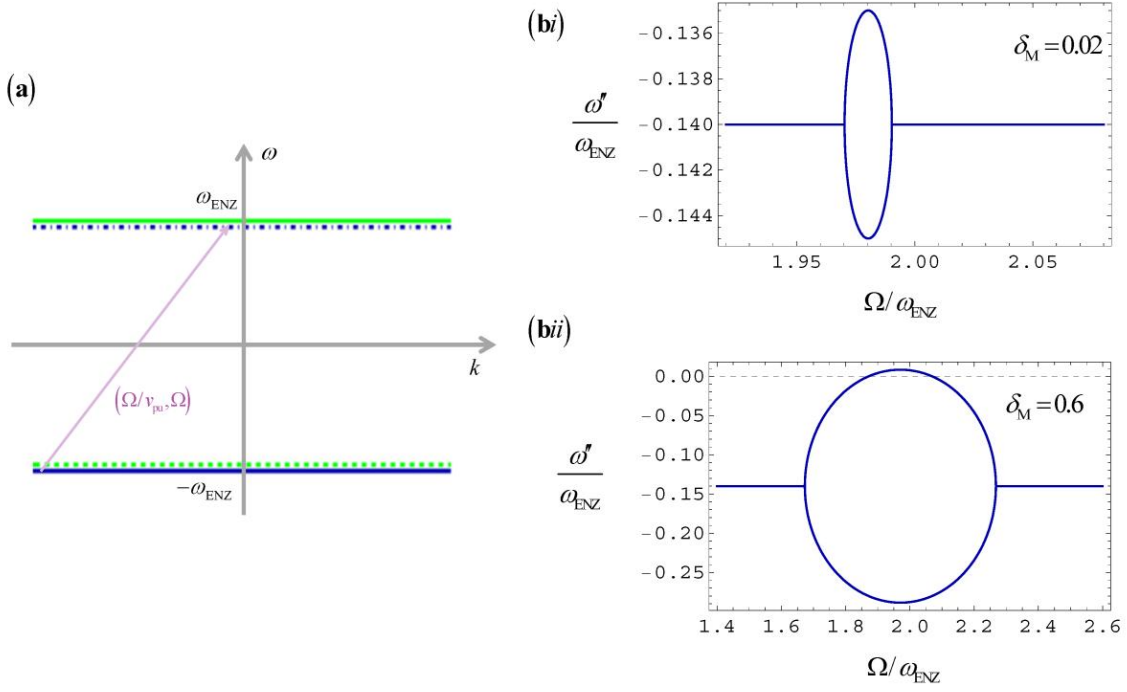

**Fig. 2 (a)** Dispersion of the longitudinal modes in the spacetime crystal for a weak modulation and $\Omega$ slightly below the parametric resonance $2\omega_{ENZ}$. The solid lines denote the static (unmodulated) dispersion, while the blue dash–dotted line (green dashed line) represents the sidebands with $l=1$ ($l=-1$). These sidebands nearly coincide with the static dispersion due to the proximity to the resonance ($\Omega \approx 2\omega_{ENZ}$). **(b)** Decay rate for the hybridized modes resulting from the coupling between the static and sideband branches for (i) weak modulation $\delta_M = 0.02$ and (ii) strong modulation $\delta_M = 0.6$. In the latter case, one of the eigenmodes exhibits a positive $\omega''$, indicating parametric amplification.

Moreover, it is straightforward to show that the normalized momentum satisfies a dissipative Mathieu equation, $\frac{\partial^2 \tilde{p}}{\partial t^2} + \Gamma \frac{\partial \tilde{p}}{\partial t} + \omega_{ENZ}^2 \left( z - v_{pu} t \right) \tilde{p} = 0$, whose solutions are well documented in the literature [49]. As previously noted, $z$ plays no dynamical role and can therefore be set to zero, recovering the standard Mathieu form. It is well established that parametric amplification can occur when the modulation frequency is approximately twice the undriven resonance frequency, i.e., when $\Omega = 2\omega_{ENZ}$. As illustrated in Fig. 2a, this parametric resonance arises when the sideband $l = +1$ of the negative-frequency branch overlaps with the static positive-frequency branch. Because the two curves coincide exactly when $\Omega = 2\omega_{ENZ}$, the parametric resonance occurs simultaneously for all *k*. Such a collective resonance has been recently discussed in the context of various plasmonic time crystals [15], and related phenomena have been analyzed in Ref. [50].

From the theory of the Mathieu equation [15, 49], the Bloch frequency ($\omega = \omega' + i\omega''$) for a modulation frequency close to the parametric resonance is approximately $\omega' \approx \omega_{ENZ}$ with the decay rate determined by $\omega'' \approx -\frac{\Gamma}{2} \pm \omega_{ENZ} \operatorname{Re}\left\{\sqrt{\left(\frac{\delta_M}{4}\right)^2 - \Delta^2}\right\}$, with $\Delta = \frac{\Omega}{2\omega_{ENZ}} - 1$ denoting the frequency detuning from resonance. Two branches emerge as a result of the hybridization between the static positive-frequency band and the negative-frequency sideband. The decay rate reaches its minimum at resonance ($\Delta = 0$): $\omega''_{max} \approx -\frac{\Gamma}{2} + \omega_{ENZ} \frac{\delta_M}{4}$.

In Figs. 2b(i) and 2b(ii), we show the exact decay rate of the longitudinal plasmons in BiTeI as a function of the normalized modulation frequency, computed using a transfer-matrix method [15]. Figure 2b(i) corresponds to the case of weak modulation ($\delta_M = 0.02$), while Fig. 2b(ii) illustrates an extremely strong modulation ($\delta_M = 0.6$). The latter scenario requires a pump field on the order of $E_{0pu} \sim 10^8 V/m$, which is not feasible under continuous-wave excitation but may be achievable using a short pulse containing only a few tens of pump cycles. As seen, for a sufficiently strong pump, it becomes possible to efficiently excite volume plasmons through anomalous velocity modulation, leading to stimulated emission of plasmon waves ($\omega''_{max} > 0$) [51]. This regime requires a modulation depth satisfying $\delta_M > \dfrac{2\Gamma}{\omega_{ENZ}} = 0.56$. Note that $\omega''_{max}$ is slightly detuned from the condition $\Delta = 0$ due to dissipative effects. While volume plasmons are typically nonradiative excitations probed through electron spectroscopy [52], the mechanism proposed here provides an alternative, all-optical pathway to access and coherently control them.

In summary, we have shown that the intrinsic $\chi^{(2)}$ nonlinearity arising from the anomalous velocity in Weyl semimetals enables the formation of optical spacetime crystals and the parametric amplification of volume plasmons. Beyond BiTeI, other low-symmetry conductors belonging to 4mm, and 6mm point groups, such as TaAs, NbAs, and NbP, exhibit even larger Berry curvature dipoles [33, 53], although their smaller bandgaps may enhance sensitivity to interband effects. Our findings reveal a fundamentally new route to ultrafast, nonreciprocal, and amplifying photonic platforms driven by quantum geometric effects.

**Acknowledgements:** This work is supported in part by the Institution of Engineering and Technology (IET), by the Simons Foundation (award SFI-MPS-EWP-00008530-10), and by national funds through FCT – Fundação para a Ciência e a Tecnologia, I.P., and, when eligible, co-funded by EU funds under project/support UID/50008/2025 – Instituto de Telecomunicações, with DOI identifier <https://doi.org/10.54499/UID/50008/2025>.

# References

[1] D. L. Sounas and A. Alù, "Non-reciprocal photonics based on time modulation," *Nat. Photon.* **11**, 774 (2017).
[2] Z. Deck-Léger, N. Chamanara, M. Skorobogatiy, M. G. Silveirinha and C. Caloz, "Uniform-velocity spacetime crystals," *Adv. Photon.* **1**, 1 – 26 (2019).
[3] C. Caloz and Z. Deck-Léger, "Spacetime metamaterials, Part I: General concepts," *IEEE Trans. Antennas Propag.* **68**, 1569 (2020).

[4] E. Galiffi, R. Tirole, S. Yin, H. Li, S. Vezzoli, P. A. Huidrobo, M. G. Silveirinha, R. Sapienza, A. Alù and J. B. Pendry, "Photonics of time-varying media," *Adv. Photon.* **4**, 014002 (2022).

[5] N. Engheta, "Four-dimensional optics using time-varying metamaterials", *Science* **379**, 1 094 (2023).

[6] M. M. Asgari, P. Garg, X. Wang, M. S. Mirmoosa, C. Rockstuhl, and V. Asadchy, "Theory and applications of photonic time crystals: a tutorial", *Adv. Opt. Photon.* **16**, 958 (2024).

[7] S. Taravati and C. Caloz, "Mixer-duplexer-antenna leaky-wave system based on periodic space-time modulation", *IEEE Trans. Antennas Propag.* **65**, 442–452 (2017).

[8] W. Jaffray, S. Saha, V. M. Shalaev, A. Boltasseva, and M. Ferrera, "Transparent conducting oxides: From all-dielectric plasmonics to a new paradigm in integrated photonics", *Adv. Opt. Photonics* **14**, 148 (2022).

[9] N. Kinsey, C. DeVault, J. Kim, M. Ferrera, V. M. Shalaev, and A. Boltasseva, "Epsilon-near-zero Al-doped ZnO for ultrafast switching at telecom wavelengths", *Optica* **2**, 616 (2015).

[10] L. Caspani *et al.*, "Enhanced nonlinear refractive index in ε-near-zero materials", *Phys. Rev. Lett.* **116**, 233901 (2016).

[11] M. Z. Alam, I. De Leon, and R. W. Boyd, "Large optical nonlinearity of indium tin oxide in its epsilon-near-zero region", *Science* **352**, 795 (2016).

[12] Y. Zhou, M. Z. Alam, M. Karimi, J. Upham, O. Reshef, C. Liu, A. E. Willner, and R. W. Boyd, "Broadband frequency translation through time refraction in an epsilon-near-zero material", *Nat. Commun.* **11**, 2180 (2020).

[13] R. Tirole, S. Vezzoli, E. Galiffi, I. Robertson, D. Maurice, B. Tilmann, S. A. Maier, J. B. Pendry, and R. Sapienza, "Double-slit time diffraction at optical frequencies", *Nat. Phys.* **19**, 999 (2023).

[14] E. Lustig, O. Segal, S. Saha, E. Bordo, S. N. Chowdhury, Y. Sharabi, A. Fleischer, A. Boltasseva, O. Cohen, V. M. Shalaev, and M. Segev, "Time-refraction optics with single cycle modulation", *Nanophotonics* **12**, 2221 (2023).

[15] J. Feinberg, D. E. Fernandes, B. Shapiro and M. G. Silveirinha, "Plasmonic Time Crystals", *Phys. Rev. Lett.*, **134**, 183801 (2025).

[16] M. Silveirinha and N. Engheta, "Tunneling of electromagnetic energy through sub-wavelength channels and bends using near-zero-epsilon materials", *Phys. Rev. Lett.* 97, 157403 (2006).

[17] J. B. Khurgin, "Photonic Time Crystals and Parametric Amplification: Similarity and Distinction," *ACS Photonics* **11**, 2150–2159 (2024).

[18] N. Konforty, M.-I. Cohen, O. Segal, Y. Plotnik, V. M. Shalaev, and M. Segev, "Second harmonic generation and nonlinear frequency conversion in photonic time-crystals", *Light: Science & Applications* **14**, 152 (2025).

[19] R. Tirole, S. Vezzoli, D. Saxena, S. Yang, T. V. Raziman, E. Galiffi, S. A. Maier, J. B. Pendry, and R. Sapienza, "Second harmonic generation at a time-varying interface", *Nat. Commun.* **15**, 7752 (2024).

[20] D. Xiao, M.-C. Chang, and Q. Niu, "Berry-phase effects on electronic properties", *Rev. Mod. Phys.* **82**, 1959–2007 (2010).

[21] D. Vanderbilt, "Berry Phases in Electronic Structure Theory: Electric Polarization, Orbital Magnetization and Topological Insulators", (Cambridge University Press, 2018).

[22] I. Sodemann and L. Fu, "Quantum nonlinear Hall effect induced by Berry curvature dipole in time-reversal invariant materials", *Phys. Rev. Lett.* **115**, 216806 (2015).

[23] Y. Onishi and L. Fu, "High-efficiency energy harvesting based on nonlinear Hall rectifier", *Phys. Rev. B* **110**, 075122 (2024)

[24] L.-K. Shi, O. Matsyshyn, J. C. W. Song, and I. S. Villadiego, "Berry-dipole photovoltaic demon and the thermodynamics of photocurrent generation within the optical gap of metals", *Phys.Rev. B* **107**, 125151 (2023).

[25] T. G. Rappoport, T. A. Morgado, S. Lannebère, and M. G. Silveirinha, "Engineering transistor-like optical gain in two-dimensional materials with Berry curvature dipoles", *Phys. Rev. Lett.* **130**, 076901 (2023).

[26] T. A. Morgado, T. G. Rappoport, S. S. Tsirkin, S. Lannebère, I. Souza, and M. G. Silveirinha, "Non-Hermitian Linear Electrooptic Effect in 3D materials", *Phys. Rev. B* **109**, 245126 (2024).

[27] A. Hakimi, K. Rouhi, T. G. Rappoport, M. G. Silveirinha, F. Capolino, "Chiral terahertz lasing with Berry curvature dipoles", *Phys. Rev. Appl.*, **22**, L041003, (2024).

[28] F. R. Prudêncio, M. G. Silveirinha, "Topological Chiral-Gain in a Berry Dipole Material", Nanophotonics, (doi.org/10.1515/nanoph-2024-0681) p. 1-13, (2025).

[29] L. Wu, S. Patankar, T. Morimoto, N. L. Nair, E. Thewalt, A. Little, J. G. Analytis, J. E. Moore, and J. Orenstein, "Giant anisotropic nonlinear optical response in transition-metal monopnictide Weyl semimetals", *Nat. Phys.* **13**, 350 (2017).

[30] Q. Fu, *et al*, "Berry curvature dipole induced giant mid-infrared second-harmonic generation in 2D Weyl semiconductor", *Adv. Mater.* **35**, 2306330 (2023).

[31] Y. Wang, J. Xiao, T.-F. Chung, Z. Nie, S. Yang, and X. Zhang, "Direct electrical modulation of second-order optical nonlinearity in a two-dimensional material", *Nat. Electron.* **4**, 725–730 (2021).

[32] S. Lannebère, N. Engheta, and M. G. Silveirinha, "Non-Hermitian Linear Electro-Optic Effect Through Interactions of Free and Bound Charges", arXiv:2503.09274 (2025).

[33] S. Lannebère, T. G. Rappoport, T. A. Morgado, I. Souza, and M. G. Silveirinha, "Symmetry Analysis of the Non-Hermitian Electro-Optic Effect in Crystals", arXiv:2502.03399 (2025).

[34] J. I. Facio, D. Efremov, K. Koepernik, J.-S. You, I. Sodemann, and J. van den Brink, "Strongly enhanced Berry dipole at topological phase transitions in BiTeI", *Phys. Rev. Lett.* **121**, 246403 (2018).

[35] C. Martin, E. D. Mun, H. Berger, V. S. Zapf, and D. B. Tanner, "Quantum oscillations and optical conductivity in Rashba spin-splitting BiTeI", *Phys. Rev. B* **87**, 041104(R) (2013).

[36] E. Cassedy and A. Oliner, "Dispersion relations in time-space periodic media: Part I—Stable interactions", *Proceedings of the IEEE* **51**, 1342 (1963).

[37] P. A. Huidobro, E. Galiffi, S. Guenneau, R. V. Craster and J. B. Pendry, "Fresnel drag in space–time-modulated metamaterials," *Proc. Natl. Acad. Sci. USA* **116**, 24943 (2019).

[38] Y. Mazor and A. Alù, "One-way hyperbolic metasurfaces based on synthetic motion", *IEEE Trans. Antennas Propag.* **68**, 1739 (2020).

[39] P. Huidobro, M. G. Silveirinha, E. Galiffi and J. B. P. Pendry, "Homogenization Theory of Space-Time Metamaterials," *Phys. Rev. Applied* **16**, 014044, (2021).

[40] F. R. Prudêncio and M. G. Silveirinha, "Replicating physical motion with Minkowskian isorefractive spacetime crystals," *Nanophotonics* **12**, 3007-3017, (2023).

[41] F. R. Prudêncio and M. G. Silveirinha, "Synthetic Axion Response with Spacetime Crystals," *Phys. Rev. Applied* **19**, 024031, (2023).

[42] F. R. Prudêncio, and M. G. Silveirinha, "Engineering Nonreciprocal Responses in Travelling-Wave Spacetime Crystals via Clausius-Mossotti Homogenization", Phys. Rev. Appl. 22, 054080, 2024.

[43] J. C. Serra, M. G. Silveirinha, "Engineering Topological Phases with a Traveling-Wave Spacetime Modulation", *Laser Photonics Rev.* (2025) (in press).

[44] J. B. Pendry, P. A. Huidobro, M. G. Silveirinha, E. Galiffi, "Crossing the light line", *Nanophotonics* **11**, 161, (2022).

[45] E. Galiffi, P. A. Huidrobo and J. B. Pendry, "An Archimedes' screw for light," *Nat. Commun.* **13**, 2523 (2022).

[46] J. C. Serra, M. G. Silveirinha, "Homogenization of Dispersive Spacetime Crystals: Anomalous Dispersion and Negative Stored Energy", *Phys. Rev. B* **108**, 035119, (2023).

[47] J. C. Serra, E. Galiffi, P. A. Huidobro, J. B. Pendry, M. G. Silveirinha, "Particle-hole instabilities in photonic time-varying systems", *Opt. Mat. Express*, **14**, 1459, (2024).

[48] A. Alex-Amor, C. Molero, M. G. Silveirinha, "Analysis of Metallic Spacetime Gratings using Lorentz Transformations", *Phys. Rev. Appl.* **20**, 014063, (2023).

[49] L. D. Landau and E. M. Lifshitz, Mechanics, 3rd ed. (Pergamon Press, Oxford, 1976), Sec. 27.

[50] X. Wang, P. Garg, M. S. Mirmoosa, A. G. Lamprianidis, C. Rockstuhl, and V. S. Asadchy, "Expanding momentum bandgaps in photonic time crystals through resonances", *Nat. Photonics* **19**, 149 (2025).

[51] D. J. Bergman, and M. I. Stockman, "Surface Plasmon Amplification by Stimulated Emission of Radiation: Quantum Generation of Coherent Surface Plasmons in Nanosystems", *Phys. Rev. Lett.* **90**, 027402 (2003).

[52] F. J. García de Abajo, "Optical excitations in electron microscopy", *Rev. Mod. Phys.* **82**, 209 (2010).

[53] Y. Zhang, Y. Sun, B. Yan, "Berry curvature dipole in Weyl semimetal materials: An ab initio study", *Phys. Rev. B* **97**, 041101(R) (2018).